\documentclass[aps,preprint,showpacs,floatfix]{revtex4}
\usepackage{bm,epsfig}
\begin{document}

\count255=\time\divide\count255 by 60 \xdef\hourmin{\number\count255}
  \multiply\count255 by-60\advance\count255 by\time
 \xdef\hourmin{\hourmin:\ifnum\count255<10 0\fi\the\count255}

\newcommand{\xbf}[1]{\mbox{\boldmath $ #1 $}}

\newcommand{\sixj}[6]{\mbox{$\left\{ \begin{array}{ccc} {#1} & {#2} &
{#3} \\ {#4} & {#5} & {#6} \end{array} \right\}$}}

\newcommand{\threej}[6]{\mbox{$\left( \begin{array}{ccc} {#1} & {#2} &
{#3} \\ {#4} & {#5} & {#6} \end{array} \right)$}}

\newcommand{\clebsch}[6]{\mbox{$\left( \begin{array}{cc|c} {#1} & {#2} &
{#3} \\ {#4} & {#5} & {#6} \end{array} \right)$}}

\newcommand{\iso}[6]{\mbox{$\left( \begin{array}{cc||c} {#1} & {#2} &
{#3} \\ {#4} & {#5} & {#6} \end{array} \right)$}}

\title{$\pi N \to$ Multi-$\pi N$ Scattering in the $1/N_c$ Expansion}

\author{Herry J. Kwee}
\email{Herry.Kwee@asu.edu}

\author{Richard F. Lebed}
\email{Richard.Lebed@asu.edu}

\affiliation{Department of Physics and Astronomy, Arizona State
University, Tempe, AZ 85287-1504}

\date{November 2006}

\begin{abstract}
We extend the $1/N_c$ expansion meson-baryon scattering formalism to
cases in which the final state contains more than two particles.  We
first show that the leading-order large $N_c$ processes proceed
through resonant intermediate states (e.g., $\rho N$ or $\pi
\Delta$).  We then tabulate linear amplitude expressions for relevant
processes and find that the pole structure of baryon resonances can be
uniquely identified by their (non)appearance in $\eta N$ or mixed
partial-wave $\pi \Delta$ final states.  We also show that
quantitative predictions of $\pi N$ to $\pi \Delta$ branching ratios
predicted at leading order alone do not agree with measurements, but
the inclusion of $1/N_c$ corrections is ample to explain the
discrepancies.
\end{abstract}

\pacs{11.15.Pg, 13.75.Gk, 13.85.Hd, 14.20.Gk}

\maketitle

\section{Introduction} \label{intro}
The baryon resonances present a peculiar conundrum in the context of
QCD\@.  On one hand, the resonances are continuum effects, often
manifesting themselves as little more than a subtle disturbance in an
otherwise smooth Argand diagram for a process such as meson-baryon
scattering or photoproduction.  On the other hand, they are states
with more or less well-defined masses, are numerous, and occur with a
regularity that suggests a spectrum delineated by multiplets of some
underlying symmetry structure~\cite{PDG}.

A growing series of
papers~\cite{CL1,CLcompat,CDLN1,CLpent,CDLN2,CLSU3,CDLM,CLpentSU3,CLSU3phenom,ItJt,CLdecouple,CLchiral}
has developed a field theory-based approach in which to tackle the
challenging problem of studying these states and the scattering
amplitudes in which they appear.  Short reviews of this literature
appear in Refs.~\cite{confs}.  The central idea rests upon symmetries
that emerge for QCD in the large $N_c$ limit, and that relate the
scattering amplitudes in channels of different $I$, $J$, and other
quantum numbers.  The original motivation dates back to a number of
papers from the 1980s that developed the group-theoretical
consequences of the Skyrme and other chiral soliton
models~\cite{ANW,HEHW,Mattis,MM}.  As one can show, the leading-order
amplitudes in the $1/N_c$ expansion, expressed in terms of $t$-channel
quantum numbers, have $I_t \! = \!  J_t$~\cite{MM}.  This is a direct
result~\cite{CL1} of unitarity in the large $N_c$ limit~\cite{DJM} and
holds for 3-flavor as well as 2-flavor processes~\cite{ItJt}.
Moreover, the result can be extended to finite-$N_c$ processes:
Amplitudes with $|I_t \! - \! J_t| \! = \! n$ are suppressed by at
least $1/N_c^n$ compared to the leading order~\cite{CDLN2,KapSavMan}.
The consequence is a systematic expansion in $1/N_c$ that can be
applied to any process involving scattering with a stable
(ground-state) baryon state, and like for any other effective theory,
a number of results hold at the lowest order in the expansion that
receive corrections from higher-order effects.  In the present case,
linear relations arise among the scattering amplitudes in different
channels, imposing degeneracies among poles (representing resonance
masses and widths) that occur within them.

Our focus here is to press beyond the baryon-plus-single meson final
state processes that were considered exclusively in all the previous
papers in this series.  A quick glance at the voluminous listings for
baryon resonances in the {\it Review of Particle Physics\/} by the
Particle Data Group (PDG)~\cite{PDG} reminds even the casual reader
that baryon resonances are common features in processes with multiple
final-state pions.  We speak here exclusively of nonstrange processes
purely as a matter of convenience; the 3-flavor formalism is more
cumbersome but is nevertheless
tractable~\cite{CLSU3,CLpentSU3,CLSU3phenom,ItJt,CLdecouple}.

\begin{figure}[ht]
\epsfxsize 1.8 in \epsfbox{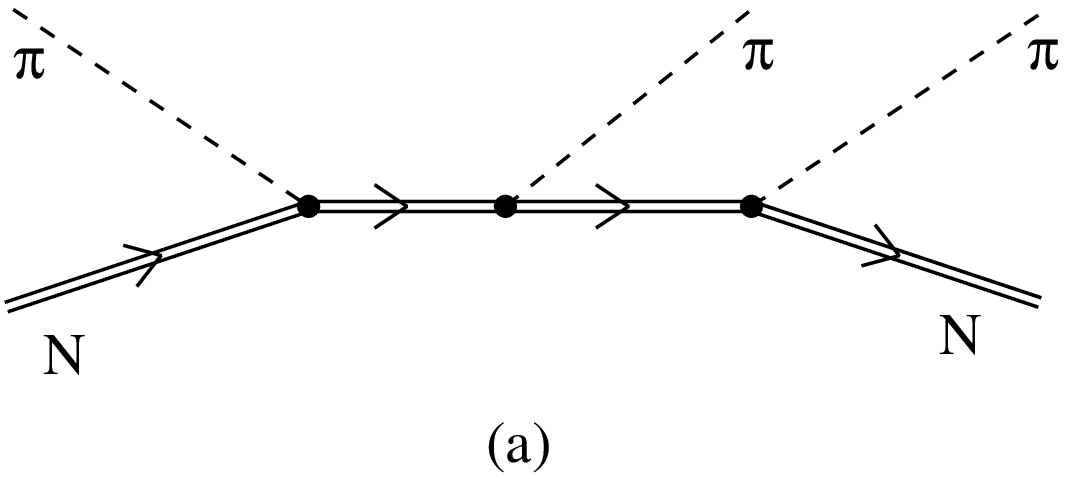} \hspace{1em}
\epsfxsize 1.8 in \epsfbox{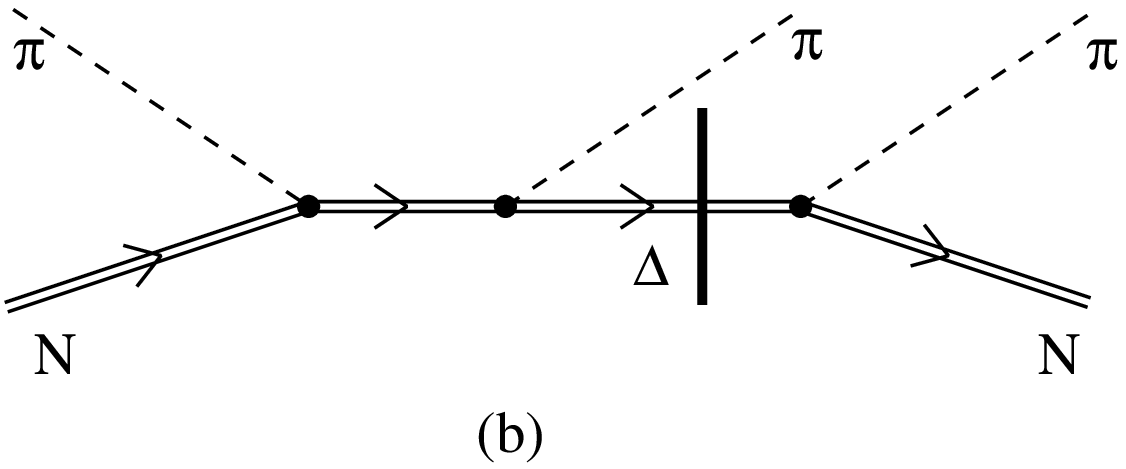} \hspace{1em}
\epsfxsize 1.8 in \epsfbox{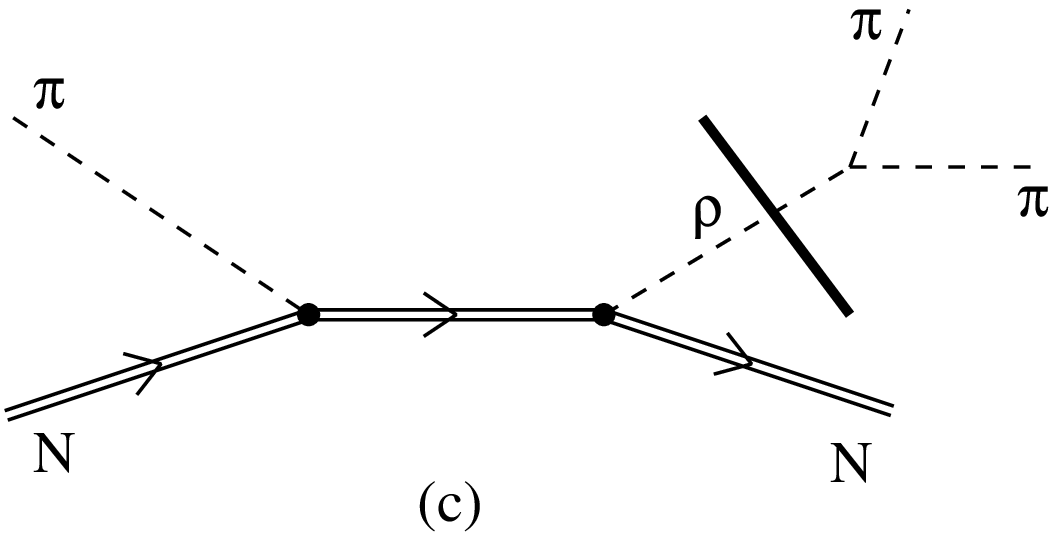}
\caption{Diagrams for $\pi N \! \to \! \pi \pi N$ scattering.  (a)
Nonresonant scattering (1 of 6 diagrams); (b) $\pi N \! \to \! \pi
\Delta$ ($\Delta$ on shell), followed by $\Delta \! \to \! \pi N$; (c)
$\pi N \! \to \! \rho N$, ($\rho$ on shell) followed by $\rho \! \to
\! \pi \pi$.}
\label{scatfigs}
\end{figure}

However, since the $1/N_c$ scattering formalism depends upon a single
incoming source and a single outgoing source scattering from the
baryon, one must employ additional considerations to find a meaningful
way to constrain such multipion processes.  In particular, standard
$N_c$ counting shows that the generic scattering amplitude for $\pi N
\! \to \! \pi N$ is $O(N_c^0)$, while that for $\pi N \! \to \! \pi
\pi N$ [Fig.~\ref{scatfigs}(a)] is $O(N_c^{-1/2})$.  This figure
represents one of the six diagrams for this processes, when all
permutations of the $\pi$-baryon vertices with respect to the external
$\pi$ states are counted; indeed, when the intermediate baryon lines
belong to the ground-state multiplet ($N$, $\Delta, \ldots$),
cancellations between the $\pi$-baryon couplings, which depend upon
the existence of this degenerate multiplet of baryon states, are
necessary to obtain the $O(N_c^{-1/2})$ scaling of the full
amplitude~\cite{DJM}.

Nevertheless, circumstances exist in which processes that {\em
eventually\/} produce two (or more) pions nevertheless appear with
amplitudes at leading order, $O(N_c^0)$.  As suggested above and to be
elucidated further in the next section, the $\Delta$ is stable for
sufficiently large $N_c$; its width scales as $1/N_c^2$.  Therefore,
as in Fig.~\ref{scatfigs}(b), the $\pi N \! \to \! \pi \pi N$ process
may be cut (indicating an on-shell state) at the intermediate stage,
$\pi N \! \to \! \pi \Delta$.  Of course, we live in the $N_c \! = \!
3$ world where $\Gamma_\Delta$ is over 100~MeV; even so,
$\Gamma_\Delta$ is considered sufficiently small (relative to its
mass) that researchers regularly extract $\pi N \! \to \! \pi \Delta$
partial widths, and the PDG sees fit to tabulate them.  In such cases,
the single source-plus-baryon scattering formalism discussed above may
be utilized.  Similarly, standard large $N_c$ counting shows all meson
widths to be $O(1/N_c^1)$; in particular, the first stage of a $\pi N
\! \to \! \pi \pi N$ process such as in Fig.~\ref{scatfigs}(c) that is
found to be dominated by an on-shell meson resonance ($\pi N \! \to \!
\rho N$ followed by $\rho \! \to \! \pi \pi$) is also $O(N_c^0)$.
Again, one may analyze such processes using the original two-body
formalism since $\pi N \! \to \! \rho N$ partial widths have been
tabulated.  As long as one is confident that the analyzed process
contains a two-body resonant intermediate state, which is tantamount
to having confidence that the continuum background of
Fig.~\ref{scatfigs}(a) is correctly subtracted from the full
experimentally measured $\pi N \! \to \pi \pi N$ scattering amplitude,
then the separation between Fig.~\ref{scatfigs}(a) and (b),(c)
processes is clean.  This applies not only to the leading-order
amplitudes, but to the $1/N_c$ subleading corrections in each case.

In this paper we tabulate all $\pi N \! \to \! \pi N$, $\pi \Delta$,
$\eta N$, $\eta \Delta$, $\rho N$, and $\omega N$ scattering
amplitudes at leading order using the large $N_c$ scattering
formalism, for all channels up to spin-5/2 for either parity ${\cal
P}$.  (Likewise, $\rho \Delta$ and $\omega \Delta$ could easily be
tabulated, but the thresholds for these states are about $2$~GeV,
where baryon resonance data becomes less abundant.)  The $\pi N$
final-state results of course represent elastic scattering, and have
been tabulated in the original papers~\cite{CL1,CLcompat}, as have
been unmixed (no change in relative orbital angular momentum) $\pi
\Delta$ amplitudes.  The older papers also presented results for the
illustrative but unphysical case of $\eta \! \to \! \eta$ scattering.
We also note that previous papers using the $1/N_c$
expansion~\cite{CGKM,CC,GSS,GS} present numerical results for excited
baryons decaying into $\eta N$ and $\Delta N$ final states, but the
assumptions required for these analyses is rather different from ours,
and we defer a direct comparison to the Conclusions.

Our approach is fraught with difficulties, both at the theoretical and
practical levels.  First, we do not explicitly include the $1/N_c$
corrections in the tabulated results.  While the previous papers have
shown how to carry out this procedure~\cite{CDLN2}, presenting all the
explicit results here would make already complicated tables positively
impenetrable.  Nevertheless, to illustrate our points about the
importance of $1/N_c$ corrections, we demonstrate in one specific case
($I \! = \! J \! = \! \frac 1 2$ positive-${\cal P}$ resonances) the
effect of these $1/N_c$ corrections.

Second, the mesons considered differ widely in mass, from 140~MeV
$\pi$ to the 783~MeV $\omega$.  We take into account differences in
phase space when making numerical comparisons, simply by using the
standard two-body formula (Eqs.~(38.16)--(38.17) of Ref.~\cite{PDG})
for $M \! \to \! m_1 m_2$.  Even so, the mesons fall into two entirely
different categories: The $\pi$ and $\eta$ are pseudo-Nambu-Goldstone
bosons of spontaneously broken chiral symmetry, while the $\rho$ and
$\omega$ are vector mesons whose masses are set by the QCD scale.
Chiral symmetry has not been imposed anywhere in this formalism
(although combining this approach and chiral symmetry has yielded some
rather interesting results~\cite{CLchiral}), and likely produce
interesting physical effects beyond those considered here.

Third, the data set, while quite extensive, is filled with internal
contradictions between one given partial-wave analysis and the next;
and even when the analyses agree, the uncertainties extracted are
often as large as the effects themselves.  This fact was an additional
motivating factor for us not to attempt to tabulate all the $O(1/N_c)$
corrections: Our opinion is that such a global analysis of all baryon
resonance observables based upon the large set of amplitudes appearing
at $O(N_c^0)$ and $O(1/N_c)$ should be carried out by the experts of
partial-wave analysis themselves, who have access to the raw data and
understand the many systematic and correlated uncertainties in the
experimental measurements.

Here we are mainly interested in identifying the underlying pole
structure of the system of nonstrange baryon resonances (classified
according to the underlying symmetry provided by large $N_c$), as
identified---for reasons to be discussed below---by the presence or
absence of certain decay channels (primarily those containing an
$\eta$ or a mixed partial-wave $\pi \Delta$ final state).  We also
find that, once identified, the pole-containing amplitudes alone do
not provide robust numerical results for the relative branching ratios
(BR's) at $O(N_c^0)$; including $O(1/N_c)$ corrections has a dramatic
effect upon the numerical predictions, and their inclusion appears to
be essential to obtain an accurate rendering of the data.  We shall
use a well-defined field theoretical expansion to peer into one of the
murkiest corners of QCD, and reveal both the successes and limitations
of the method.

This paper is organized as follows: In Sec.~\ref{group} we present the
explicit expressions used to derive the leading-order [$O(N_c^0)$]
scattering amplitudes for all meson-baryon to meson-baryon scattering
amplitudes.  In Sec.~\ref{result} we tabulate all $\pi N \! \to \! \pi
N$, $\pi \Delta$, $\eta N$, $\eta \Delta$, $\rho N$, and $\omega N$
scattering amplitudes, for all channels up to spin-5/2 for either
${\cal P}$.  Section~\ref{phenom} presents an analysis, channel by
channel, of results from comparing existing resonance data to our
results and consider the evidence for resonant multiplets.  In
Sec.~\ref{concl} we present a brief discussion (including a comparison
to related previous large $N_c$ work) and conclude.

\section{Group Theory Preliminaries} \label{group}

We begin with a brief description of the linear expressions describing
meson-baryon scattering amplitudes and degeneracies among the baryonic
resonances embedded within them.  The formalism from which they are
derived~\cite{HEHW,MM,Mattis} is obtained from considering only
initial- and final-state baryons lying in the ground-state band of
large $N_c$, which is the completely symmetric spin-flavor
representation generalizing the {\bf 56} of SU(6), and whose nonstrange
members have spin = isospin $R \! = \! \frac 1 2, \frac 3 2, \ldots,
\frac{N_c}{2}$.  The $\Delta$ is therefore, like the nucleon, a stable
baryon in the large $N_c$ limit [$m_\Delta \! - \! m_N \! \equiv
\delta \! = \!  O(1/N_c)$] and decays in the case of $N_c \!  = \! 3$
only because the chiral limit is approached more quickly than the
large $N_c$ limit: $m_\pi \! < \! \delta_{N_c = 3}$.  But the
restriction of baryons to this multiplet is not physically
constraining, since all observed meson-baryon scattering processes fit
into this category.  This analysis can in principle be carried out for
baryons (in the ground-state multiplet) and mesons of arbitrary spin,
isospin, and strangeness quantum numbers, although in the context of
this paper we examine only processes created by $\pi N$ scattering,
and restrict the final states to consist of nonstrange mesons and
baryons.  The basic process is
\begin{equation}
m + B \to m^\prime + B^\prime ,
\end{equation}
where $m \, (m^\prime)$ is a meson of spin $s \, (s^\prime)$ and
isospin $i \, (i^\prime)$, in a state of relative orbital angular
momentum $L$ ($L^\prime$) with a baryon $B \, (B^\prime)$ of spin =
isospin $R \, (R^\prime)$ in the ground-state multiplet, and the
total spin angular momentum (not including relative orbital angular
momentum) of the meson and baryon is denoted $S \, (S^\prime)$.  The
intermediate state is labeled by the total quantum numbers $I$ and
$J$, giving the full partial wave $S_{LL^\prime SS^\prime IJ}$, where
the meson and baryon quantum numbers are implicit.  Abbreviating the
multiplicity $2X\!+\!1$ of an SU(2) representation of quantum number
$X$ by $[X]$, then one finds~\cite{MM,CLcompat}
\begin{eqnarray}
S_{L L^\prime S S^\prime I J} & = & \sum_{K, \tilde{K} ,
\tilde{K}^\prime} [K]
([R][R^\prime][S][S^\prime][\tilde{K}][\tilde{K}^\prime])^{1/2}
\nonumber \\
& & \times \left\{ \begin{array}{ccc}
L & i & \tilde{K} \\
S & R & s \\
J & I & K \end{array} \right\}
\left\{ \begin{array}{ccc}
L^\prime & i^\prime & \tilde{K}^\prime \\
S^\prime & R^\prime & s^\prime \\
J & I & K \end{array} \right\}
\tau_{K \tilde{K} \tilde{K}^\prime L L^\prime} . \label{Mmaster}
\end{eqnarray}
The remaining symbols, $K$, $\tilde{K}$, and $\tilde{K}^\prime$, are
intermediate quantum numbers: In generalizing the process to allow for
mesons $m$, $m^\prime$ of arbitrary isospin and spin, one requires the
quantum numbers ${\bf K} \! \equiv \! {\bf I} \! + \!  {\bf J}$,
$\tilde {\bf K} \! \equiv \! {\bf i} \! + \! {\bf L}$, and $\tilde
{\bf K}^\prime \! \equiv \!  {\bf i^\prime} \!  + \!  {\bf L}^\prime$
(so that ${\bf K} \! = \! \tilde {\bf K} \! + \!  {\bf s}
\! = \! \tilde {\bf K}^\prime \!  + \! {\bf s^\prime}$) used in
Eq.~(\ref{Mmaster}).  $S_{LL^\prime S S^\prime IJ}$ is the (isospin-
and angular momentum-reduced) $S$ matrix for this channel reduced in
the sense of the Wigner-Eckart theorem, the factors in braces are $9j$
coefficients, and $\tau_{K\tilde{K}\tilde{K}^\prime LL^\prime}$ are
universal amplitudes ({\it reduced\/} or {\it K\/} amplitudes) that
are independent of $I$, $J$, $R$, $R'$, $i$, $i^\prime$, $s$, and
$s^\prime$.  The linear relations among the scattering amplitudes can
be seen from the structure of Eq.~(\ref{Mmaster}); the point is simply
that there are more $S_{LL^\prime S S^\prime IJ}$ amplitudes than
there are $\tau_{K\tilde{K}\tilde{K}^\prime LL^\prime}$
amplitudes. Thus, at leading order in $1/N_c$ one finds linear
constraints between the $S_{LL^\prime S S^\prime IJ}$ partial waves.
Moreover, in the case of a spinless meson ($s \! = \! 0$ or $s^\prime
\! = \! 0$) or an isoscalar meson ($i \! = \! 0$ or $i^\prime \! = \!
0$), the $9j$ symbol containing the relevant zero collapses to a $6j$
symbol, or even (if a single $9j$ symbol contains two zeroes) to a
trivial closed-form result, radically simplifying the result.

Equation~(\ref{Mmaster}) was first derived in the context of the
Skyrme model~\cite{HEHW,MM,Mattis}.  In this picture, as in other
chiral soliton models, the ``hedgehog configuration'' soliton at the
classical or mean-field level (which dominates as $N_c \! \rightarrow
\! \infty)$ breaks both rotational and isospin symmetries but is
invariant under ${\bf K} \equiv {\bf I}+ {\bf J}$.  Accordingly, the
intrinsic dynamics of the soliton commutes with the ``grand spin'' of
${\bf I} \! + \! {\bf J}$, and excitations can be labeled by $K$,
which is the $K$ of Eq.~(\ref{Mmaster}).  Note that the physical
states are projected from the hedgehogs, so that ${\bf K}$ of the
physical state is {\em not\/} just ${\bf I} \! + \! {\bf J}$, but
rather represents the grand spin of one underlying intrinsic state;
the physical state is a linear combination of hedgehog states that in
total has good $I$ and $J$.

The derivation of Eq.~(\ref{Mmaster}) from chiral soliton models has
the advantage of suggesting a clear physical picture in which the $K$
quantum number has a simple interpretation.  Of course, it has the
disadvantage of being based on a model rather than directly on large
$N_c$ QCD.  However, these relations are in fact exact results in the
large $N_c$ limit of QCD and do not depend upon any additional model
assumptions.  A direct derivation for the spinless meson case (not an
essential restriction) based upon large $N_c$ consistency
rules~\cite{DJM} and exploiting the famous $I_t \! = \! J_t$
rule~\cite{MM} is given in the appendix of Ref.~\cite{CL1}.

Multiplets of baryon resonances degenerate in both mass and width
appear immediately from the structure of Eq.~(\ref{Mmaster}), provided
one defines the resonance position in scattering amplitudes to be at
the pole.  In order for one of the $S_{LL^\prime S S^\prime IJ}$
amplitudes to contain such a pole, one of the
$\tau_{K\tilde{K}\tilde{K}^\prime LL^\prime}$ amplitudes must contain
the pole.  However, since the $\tau_{K\tilde{K}\tilde{K}^\prime
LL^\prime}$ amplitudes each contribute to multiple partial waves, all
of these channels must have resonances at the same position, which
means that the masses and widths of certain resonances in different
partial waves must be degenerate.  Moreover, while the amplitudes
themselves, as functions of energy, carry dependence on the orbital
angular momenta $L$, $L^\prime$ and the auxiliary variables
$\tilde{K}$, $\tilde{K}^\prime$, the resonances themselves (once
formed) have no memory of the process used to create them.  Resonant
poles are therefore poles classified solely by the underlying quantum
number $K$, a fact we freely exploit in our analysis.

\section{Amplitude Tables} \label{result}

The results of transition amplitude calculations for $\eta N$, $\eta
\Delta$, $\pi N$, $\pi \Delta$, $\omega N$, and $\rho N$ final states are
collected for spins $\frac 1 2$, $\frac 3 2$, and $\frac 5 2$ in
Tables~\ref{t1}--\ref{t6}.  Tables~\ref{t1} and~\ref{t2} contain
results for $I \! = \! \frac 1 2$ positive-${\cal P}$ resonances,
Tables~\ref{t3} and~\ref{t4} for $I \! = \! \frac 1 2$ negative-${\cal
P}$ resonances, Table~\ref{t5} for $I \! = \! \frac 3 2$
positive-${\cal P}$ resonances and Table~\ref{t6} for $I \! = \! \frac
3 2$ negative-${\cal P}$ resonances.  Since ${\cal P}$ is a good
symmetry of strong interactions, the set of reduced amplitudes for
each ${\cal P}$ are distinct; working only in the context of this
formalism, one has no reason to expect degeneracies between poles
carrying opposite ${\cal P}$.  For each of these states, we present
all the possible partial-wave amplitudes with the final-state total
spins separately specified (as can be separated via helicity amplitude
decomposition) when more than one possibility occurs.  Thus, the most
general notation needed for $\pi N \! \to \! m^\prime B^\prime$
partial waves is $LL^{\prime \; (\pi N) (m^\prime
B^\prime)_{S^\prime}}_{2I, 2J}$; if $L^\prime \! = \!  L$ then the
label $L^\prime$ is suppressed, while if $s^\prime$ (the $m^\prime$
spin) is zero, then $S^\prime$ equals the spin of baryon $B^\prime$
and is suppressed.

The results for $\pi N$ and $L \! = \! L^\prime$ $\pi \Delta$ final
states have been tabulated previously~\cite{CL1,CLcompat}, as becomes
clear upon noting the simplification~\cite{CLcompat} for spinless
$\pi$'s: $s^\pi_{KLL^\prime} \!  = \! (-1)^{L-L^\prime}
\tau_{KKKLL^\prime}$.

In total, the tables present results for 92 measurable partial-wave
amplitudes, and are written in terms of 35 distinct reduced amplitudes
that appear at the leading ($N_c^0$) order.  Had the possible but not
yet phenomenologically interesting $\rho \Delta$ and $\omega \Delta$
channels been included, the degree of degeneracy would have been even
more pronounced.  Included in this total of 35 is the effect of the
time-reversal invariance constraint $\tau_{K\tilde{K}\tilde{K}^\prime
LL^\prime} \!  = \! \tau_{K\tilde{K}^\prime\tilde{K} L^\prime L}$,
which is apparent when one considers the physical origin of the
various indices.  If $O(1/N_c^1)$ amplitudes, satisfying the
constraint $|I_t \! - \! J_t| \! = \! 1$ are included (the sort of
analysis performed in Ref.~\cite{CDLN2}), the total would perhaps
triple, in which case the physical partial waves and reduced
amplitudes would be roughly comparable in number---although
Ref.~\cite{CDLN2} shows that at least a few ``gold-plated'' relations
[ones that include the $O(1/N_c)$ corrections] would survive.

\begin{table}
\caption{Partial-wave amplitudes for positive-parity $N$ resonances in
multipion processes (the $\pi N$ final state is included for
comparison).  Expansions are given in terms of $K$ amplitudes.
\label{t1}}
\begin{tabular}{lcccccl}
State \mbox{  } && Poles \mbox{   } &&
\multicolumn{3}{l}{Partial Wave, $K$-Amplitudes} \\
\hline\hline
$N_{1/2}^{+}$ && $K = 0, 1$ &&
$P^{{(\pi N)} {(\eta N)}}_{11}$
&=& $-\frac{\sqrt 2}{\sqrt3}\tau_{11111}$ \\
&& && $P^{{(\pi N)} {(\pi N)}}_{11}$
&=& $\frac{1}{3}\tau_{00011} + \frac{2}{3}\tau_{11111}$ \\
&& && $P^{{(\pi N)} {(\pi \Delta)}}_{11}$
&=& $\frac{\sqrt 2}{3}\tau_{00011} - \frac{\sqrt 2}{3} \tau_{11111}$ \\
&& && $P^{{(\pi N)} {(\omega N)}_1}_{11}$
&=& $\frac{1}{3}\tau_{00111} +\frac{2}{3} \tau_{11111}$ \\
&& && $P^{{(\pi N)} {(\omega N)}_3}_{11}$
&=& $\frac{\sqrt 2}{3}\tau_{00111} -\frac{\sqrt 2}{3} \tau_{11111}$ \\
&& && $P^{{(\pi N)} {(\rho N)}_1}_{11}$
&=& $\frac{\sqrt 2}{3\sqrt 3}\tau_{00111} - \frac{\sqrt 2}{9} \tau_{11011} +
\frac{2\sqrt{10}}{9}\tau_{11211}$\\
&& && $P^{{(\pi N)} {(\rho N)}_3}_{11}$
&=& $-\frac{1}{3\sqrt 3}\tau_{00111} - \frac{4}{9}\tau_{11011}$ \\
&& && &&$+\frac{1}{\sqrt 3}\tau_{11111} + \frac{\sqrt 5}{9}\tau_{11211}$ \\
\hline
$N_{3/2}^{+}$ && $K = 1, 2$ &&
$P^{{(\pi N)} {(\eta N)}}_{13}$
&=& $\frac{1}{\sqrt 6}\tau_{11111}$ \\
&& && $P^{{(\pi N)} {(\pi N)}}_{13}$
&=& $\frac{1}{6}\tau_{11111} + \frac{5}{6}\tau_{22211}$ \\
&& && $P^{{(\pi N)} {(\pi \Delta)}}_{13}$
&=& $\frac{\sqrt 5}{6}\tau_{11111} - \frac{\sqrt 5}{6} \tau_{22211}$ \\
&& && $PF^{{(\pi N)} {(\pi \Delta)}}_{13}$
&=& $-\frac{\sqrt 5}{\sqrt 6}\tau_{22213}$ \\
&& && $P^{{(\pi N)} {(\omega N)}_1}_{13}$
&=& $\frac{1}{6}\tau_{11111} + \frac{5}{6}\tau_{22111}$ \\
&& && $P^{{(\pi N)} {(\omega N)}_3}_{13}$
&=& $\frac{\sqrt 5}{6}\tau_{11111} - \frac{\sqrt 5}{6} \tau_{22111}$ \\
&& && $PF^{{(\pi N)} {(\omega N)}_3}_{13}$
&=& $-\frac{\sqrt 5}{\sqrt 6}\tau_{22313}$ \\
&& && $P^{{(\pi N)} {(\rho N)}_1}_{13}$
&=& $-\frac{\sqrt 2}{9} \tau_{11011} +\frac{1}{2\sqrt 6}\tau_{11111}
-\frac{\sqrt 5}{18\sqrt 2}\tau_{11211}$ \\
&& && && $-\frac{5}{6\sqrt 6}\tau_{22111}
+\frac{5}{6\sqrt 2}\tau_{22211}$\\
&& && $P^{{(\pi N)} {(\rho N)}_3}_{13}$
&=& $\frac{\sqrt 5}{9\sqrt 2}\tau_{11011} -\frac{1}{9\sqrt 2}\tau_{11211}$ \\
&& && && $+\frac{\sqrt{10}}{3\sqrt 3}\tau_{22111}
+\frac{\sqrt 5}{3\sqrt 2}\tau_{22211}$  \\
&& && $PF^{{(\pi N)} {(\rho N)}_3}_{13}$
&=& $\frac{1}{2\sqrt 3}\tau_{11213} +\frac{\sqrt 5}{6\sqrt 3}\tau_{22213}
+\frac{\sqrt{10}}{3\sqrt 3}\tau_{22313}$ \\
\hline
\end{tabular}
\end{table}

\begin{table}
\caption{First continuation of Table~\ref{t1}.\label{t2}}
\begin{tabular}{lcccccl}
State \mbox{  } && Poles \mbox{   } &&
\multicolumn{3}{l}{Partial Wave, $K$-Amplitudes} \\
\hline\hline
$N_{5/2}^{+}$ && $K = 2, 3$ &&
$F^{{(\pi N)} {(\eta N)}}_{15}$
&=& $-\frac{2}{3}\tau_{33333}$ \\
&& && $F^{{(\pi N)} {(\pi N)}}_{15}$
&=& $\frac{5}{9}\tau_{22233} +\frac{4}{9}\tau_{33333}$ \\
&& && $FP^{{(\pi N)} {(\pi \Delta)}}_{15}$
&=& $\frac{\sqrt 5}{3}\tau_{22231}$ \\
&& && $F^{{(\pi N)} {(\pi \Delta)}}_{15}$
&=& $\frac{2\sqrt 5}{9}\tau_{22233} -\frac{2\sqrt 5}{9} \tau_{33333}$ \\
&& && $F^{{(\pi N)} {(\omega N)}_1}_{15}$
&=& $\frac{5}{9}\tau_{22333} +\frac{4}{9}\tau_{33333}$ \\
&& && $FP^{{(\pi N)} {(\omega N)}_3}_{15}$
&=& $\frac{\sqrt 5}{3}\tau_{22131}$ \\
&& && $F^{{(\pi N)} {(\omega N)}_3}_{15}$
&=& $\frac{2\sqrt 5}{9}\tau_{22333} -\frac{2\sqrt 5}{9} \tau_{33333}$ \\
&& && $F^{{(\pi N)} {(\rho N)}_1}_{15}$
&=& $-\frac{5\sqrt 2}{27}\tau_{22233} +\frac{10}{27}\tau_{22333}
-\frac{2\sqrt 5}{27\sqrt 7}\tau_{33233}$ \\
&& && && $-\frac{4}{27}\tau_{33333}
+\frac{4}{3\sqrt 7}\tau_{33433}$\\
&& && $FP^{{(\pi N)} {(\rho N)}_3}_{15}$
&=& $\frac{\sqrt 5}{3\sqrt 6}\tau_{22131}
-\frac{\sqrt 5}{9\sqrt 2}\tau_{22231} +\frac{2\sqrt 7}{9}\tau_{33231}$ \\
&& && $F^{{(\pi N)} {(\rho N)}_3}_{15}$
&=& $\frac{4\sqrt{10}}{27}\tau_{22233}
+\frac{\sqrt 5}{27}\tau_{22333} +\frac{8}{27\sqrt 7}\tau_{33233}$ \\
&& && && $+\frac{5\sqrt 5}{27}\tau_{33333}
+\frac{\sqrt 5}{3\sqrt 7}\tau_{33433}$ \\
\hline
\end{tabular}
\end{table}

\begin{table}
\caption{Partial-wave amplitudes for negative-parity $N$ resonances in
multipion processes (the $\pi N$ final state is included for
comparison). Expansions are given in terms of $K$ amplitudes.
\label{t3}}
\begin{tabular}{lcccccl}
State \mbox{  } && Poles \mbox{   } &&
\multicolumn{3}{l}{Partial Wave, $K$-Amplitudes} \\
\hline\hline
$N_{1/2}^{-}$ && $K = 1$ &&
$S^{{(\pi N)} {(\eta N)}}_{11}$
&=& $0$ \\
&& && $S^{{(\pi N)} {(\pi N)}}_{11}$
&=& $\tau_{11100}$ \\
&& && $SD^{{(\pi N)} {(\pi \Delta)}}_{11}$
&=& $- \tau_{11102}$ \\
&& && $S^{{(\pi N)} {(\omega N)}_1}_{11}$
&=& $\tau_{11000}$ \\
&& && $SD^{{(\pi N)} {(\omega N)}_3}_{11}$
&=& $- \tau_{11202}$ \\
&& && $S^{{(\pi N)} {(\rho N)}_1}_{11}$
&=& $\sqrt{\frac{2}{3}}\tau_{11100}$\\
&& && $SD^{{(\pi N)} {(\rho N)}_3}_{11}$
&=& $\frac{1}{\sqrt 6}\tau_{11102} + \frac{1}{\sqrt 2}\tau_{11202}$ \\
\hline
$N_{3/2}^{-}$ && $K = 1, 2$ &&
$D^{{(\pi N)} {(\eta N)}}_{13}$
&=& $-\frac{1}{\sqrt 2}\tau_{22222}$ \\
&& && $D^{{(\pi N)} {(\pi N)}}_{13}$
&=& $\frac{1}{2}\tau_{11122} + \frac{1}{2}\tau_{22222}$ \\
&& && $DS^{{(\pi N)} {(\pi \Delta)}}_{13}$
&=& $\frac{1}{\sqrt 2}\tau_{11120}$ \\
&& && $D^{{(\pi N)} {(\pi \Delta)}}_{13}$
&=& $\frac{1}{2}\tau_{11122} - \frac{1}{2} \tau_{22222}$ \\
&& && $D^{{(\pi N)} {(\omega N)}_1}_{13}$
&=& $\frac{1}{2}\tau_{11222} +\frac{1}{2}\tau_{22222}$ \\
&& && $DS^{{(\pi N)} {(\omega N)}_3}_{13}$
&=& $\frac{1}{\sqrt 2}\tau_{11020}$ \\
&& && $D^{{(\pi N)} {(\omega N)}_3}_{13}$
&=& $\frac{1}{2}\tau_{11222} -\frac{1}{2} \tau_{22222}$ \\
&& && $D^{{(\pi N)} {(\rho N)}_1}_{13}$
&=& $-\frac{1}{2\sqrt 6}\tau_{11122} +\frac{1}{2\sqrt 2}\tau_{11222}
-\frac{1}{2\sqrt{30}}\tau_{22122}$ \\
&& && && $-\frac{1}{6\sqrt 2}\tau_{22222}
+\frac{\sqrt{14}}{3\sqrt 5}\tau_{22322}$\\
&& && $DS^{{(\pi N)} {(\rho N)}_3}_{13}$
&=& $-\frac{1}{2\sqrt 3}\tau_{11120} +\frac{\sqrt 5}{2\sqrt
3}\tau_{22120}$ \\
&& && $D^{{(\pi N)} {(\rho N)}_3}_{13}$
&=& $\frac{1}{\sqrt 6}\tau_{11122} +\frac{1}{\sqrt{30}}\tau_{22122}$ \\
&& && && $+\frac{\sqrt 2}{3}\tau_{22222} +\frac{\sqrt{7}}{3\sqrt{10}}\tau_{22322}$ \\
\hline
\end{tabular}
\end{table}

\begin{table}
\caption{First continuation of Table~\ref{t3}.\label{t4}}
\begin{tabular}{lcccccl}
State \mbox{  } && Poles \mbox{   } &&
\multicolumn{3}{l}{Partial Wave, $K$-Amplitudes} \\
\hline\hline
$N_{5/2}^{-}$ && $K = 2, 3$ &&
$D^{{(\pi N)} {(\eta N)}}_{15}$
&=& $\frac{\sqrt 2}{3}\tau_{22222}$ \\
&& && $D^{{(\pi N)} {(\pi N)}}_{15}$
&=& $\frac{2}{9}\tau_{22222} + \frac{7}{9}\tau_{33322}$ \\
&& && $D^{{(\pi N)} {(\pi \Delta)}}_{15}$
&=& $\frac{\sqrt{14}}{9}\tau_{22222} - \frac{\sqrt{14}}{9} \tau_{33322}$ \\
&& && $DG^{{(\pi N)} {(\pi \Delta)}}_{15}$
&=& $-\frac{\sqrt 7}{3}\tau_{33324}$ \\
&& && $D^{{(\pi N)} {(\omega N)}_1}_{15}$
&=& $\frac{2}{9}\tau_{22222} + \frac{7}{9}\tau_{33222}$ \\
&& && $D^{{(\pi N)} {(\omega N)}_3}_{15}$
&=& $\frac{\sqrt{14}}{9}\tau_{22222} - \frac{\sqrt{14}}{9} \tau_{33222}$ \\
&& && $DG^{{(\pi N)} {(\omega N)}_3}_{15}$
&=& $-\frac{\sqrt 7}{3}\tau_{33424}$ \\
&& && $D^{{(\pi N)} {(\rho N)}_1}_{15}$
&=& $-\frac{2\sqrt 2}{3\sqrt{15}}\tau_{22122} +\frac{4\sqrt 2}{27}\tau_{22222}
-\frac{\sqrt{14}}{27\sqrt 5}\tau_{22322}$ \\
&& && && $-\frac{7\sqrt 2}{27}\tau_{33222}
+\frac{14}{27}\tau_{33322}$\\
&& && $D^{{(\pi N)} {(\rho N)}_3}_{15}$
&=& $\frac{\sqrt 7}{3\sqrt{15}}\tau_{22122} +\frac{\sqrt 7}{9}\tau_{22222}
-\frac{4}{27\sqrt 5}\tau_{22322}$ \\
&& && && $+\frac{5\sqrt 7}{27}\tau_{33222}
+\frac{4\sqrt{14}}{27}\tau_{33322}$ \\
&& && $DG^{{(\pi N)} {(\rho N)}_3}_{15}$
&=& $\frac{\sqrt{10}}{9}\tau_{22324} +\frac{\sqrt 7}{18}\tau_{33324}
+\frac{\sqrt{35}}{6\sqrt 3}\tau_{33424}$ \\
\hline
\end{tabular}
\end{table}

\begin{table}
\caption{Partial-wave amplitudes for positive-parity $\Delta$ resonances
in multipion processes (the $\pi N$ final state is included for
comparison).  Expansions are given in terms of $K$ amplitudes.
\label{t5}}
\begin{tabular}{lcccccl}
State \mbox{  } && Poles \mbox{   } &&
\multicolumn{3}{l}{Partial Wave, $K$-Amplitudes} \\
\hline\hline
$\Delta_{1/2}^{+}$ && $K = 1, 2$ &&
$P^{{(\pi N)} {(\eta \Delta)}}_{31}$
&=& $-\frac{1}{\sqrt 6}\tau_{11111}$ \\
&& && $P^{{(\pi N)} {(\pi N)}}_{31}$
&=& $\frac{1}{6}\tau_{11111} + \frac{5}{6}\tau_{22211}$ \\
&& && $P^{{(\pi N)} {(\pi \Delta)}}_{31}$
&=& $\frac{\sqrt 5}{6}\tau_{11111} - \frac{\sqrt 5}{6} \tau_{22211}$ \\
&& && $P^{{(\pi N)} {(\rho N)}_1}_{31}$
&=& $-\frac{\sqrt 2}{9} \tau_{11011} -\frac{1}{2\sqrt 6}\tau_{11111}
-\frac{\sqrt 5}{18\sqrt 2}\tau_{11211}$ \\
&& && && $-\frac{5}{6\sqrt 6}\tau_{22111}
-\frac{5}{6\sqrt 2}\tau_{22211}$\\
&& && $P^{{(\pi N)} {(\rho N)}_3}_{31}$
&=& $-\frac{1}{18}\tau_{11011} -\frac{1}{4\sqrt 3}\tau_{11111}
-\frac{5\sqrt 5}{36}\tau_{11211}$ \\
&& && && $+\frac{5}{12\sqrt 3}\tau_{22111}
+\frac{5}{12}\tau_{22211}$ \\
\hline
$\Delta_{3/2}^{+}$ && $K = 0, 1, 2$ &&
$P^{{(\pi N)} {(\eta \Delta)}}_{33}$
&=& $-\frac{\sqrt 5}{2\sqrt 3}\tau_{11111}$ \\
&& && $P^{{(\pi N)} {(\pi N)}}_{33}$
&=& $\frac{1}{6}\tau_{00011} + \frac{5}{12}\tau_{11111}
+\frac{5}{12}\tau_{22211}$ \\
&& && $P^{{(\pi N)} {(\pi \Delta)}}_{33}$
&=& $\frac{1}{3\sqrt 2}\tau_{00011} +\frac{1}{3\sqrt 2}\tau_{11111} -
\frac{\sqrt2}{3} \tau_{22211}$ \\
&& && $PF^{{(\pi N)} {(\pi \Delta)}}_{33}$
&=& $\frac{1}{2\sqrt 3}\tau_{22213}$ \\
&& && $P^{{(\pi N)} {(\rho N)}_1}_{33}$
&=& $-\frac{1}{6\sqrt 6}\tau_{00111} +\frac{5}{18\sqrt 2}\tau_{11011}$ \\
&& && && $-\frac{\sqrt 5}{18\sqrt 2}\tau_{11211}
+\frac{5}{6\sqrt 6}\tau_{22111}$\\
&& && $P^{{(\pi N)} {(\rho N)}_3}_{33}$
&=& $-\frac{\sqrt 5}{6\sqrt 6}\tau_{00111} +\frac{\sqrt 5}{9\sqrt
2}\tau_{11011} +\frac{\sqrt 5}{4\sqrt 6}\tau_{11111}$ \\
&& && && $-\frac{13}{36\sqrt 2}\tau_{11211} +\frac{\sqrt 5}{12\sqrt
6}\tau_{22111} +\frac{\sqrt 5}{4\sqrt 2}\tau_{22211}$ \\
&& && $PF^{{(\pi N)} {(\rho N)}_3}_{33}$
&=& $-\frac{1}{4\sqrt 3}\tau_{11213} -\frac{\sqrt 5}{12\sqrt
3}\tau_{22213} -\frac{\sqrt 5}{3\sqrt 6}\tau_{22313}$ \\
\hline
$\Delta_{5/2}^{+}$ && $K = 2, 3, 4$ &&
$F^{{(\pi N)} {(\eta \Delta)}}_{35}$
&=& $-\frac{\sqrt 5}{3\sqrt 2}\tau_{33333}$ \\
&& && $F^{{(\pi N)} {(\pi N)}}_{35}$
&=& $\frac{5}{63}\tau_{22233} + \frac{5}{18}\tau_{33333}
+\frac{9}{14}\tau_{44433}$ \\
&& && $FP^{{(\pi N)} {(\pi \Delta)}}_{35}$
&=& $-\frac{1}{3\sqrt 2}\tau_{22231}$ \\
&& && $F^{{(\pi N)} {(\pi \Delta)}}_{35}$
&=& $\frac{8\sqrt 2}{63}\tau_{22233} +\frac{7}{18\sqrt 2}\tau_{33333}
-\frac{9}{14\sqrt2}\tau_{44433}$ \\
&& && $F^{{(\pi N)} {(\rho N)}_1}_{35}$
&=& $-\frac{25}{189\sqrt 2}\tau_{22233} -\frac{20}{189}\tau_{22333}
-\frac{2\sqrt 5}{27\sqrt 7}\tau_{33233}$ \\
&& && && $-\frac{25}{108}\tau_{33333}
-\frac{5}{12\sqrt 7}\tau_{33433}$\\
&& && && $-\frac{3}{28}\tau_{44333}
-\frac{3\sqrt{15}}{28}\tau_{44433}$\\
&& && $FP^{{(\pi N)} {(\rho N)}_3}_{35}$
&=& $-\frac{\sqrt 5}{6\sqrt 6}\tau_{22131} +\frac{\sqrt 5}{18\sqrt
2}\tau_{22231} -\frac{\sqrt 7}{9}\tau_{33231}$ \\
&& && $F^{{(\pi N)} {(\rho N)}_3}_{35}$
&=& $\frac{\sqrt{10}}{189}\tau_{22233} -\frac{11\sqrt 5}{189}\tau_{22333}
+\frac{\sqrt 7}{54}\tau_{33233}$ \\
&& && && $+\frac{13\sqrt 5}{216}\tau_{33333}
-\frac{\sqrt{35}}{24}\tau_{33433}$ \\
&& && && $+\frac{3\sqrt 5}{56}\tau_{44333} +\frac{15\sqrt 3}{56}\tau_{44433}$
\\
\hline
\end{tabular}
\end{table}

\begin{table}
\caption{Partial-wave amplitudes for negative-parity $\Delta$ resonances
in multipion processes (the $\pi N$ final state is included for
comparison).  Expansions are given in terms of $K$ amplitudes.
\label{t6}}
\begin{tabular}{lcccccl}
State \mbox{  } && Poles \mbox{   } &&
\multicolumn{3}{l}{Partial Wave, $K$-Amplitudes} \\
\hline\hline
$\Delta_{1/2}^{-}$ && $K = 1$ &&
$S^{{(\pi N)} {(\eta \Delta)}}_{31}$
&=& $0$ \\
&& && $S^{{(\pi N)} {(\pi N)}}_{31}$
&=& $\tau_{11100}$ \\
&& && $SD^{{(\pi N)} {(\pi \Delta)}}_{31}$
&=& $\frac{1}{\sqrt{10}}\tau_{11102}$ \\
&& && $S^{{(\pi N)} {(\rho N)}_1}_{31}$
&=& $-\frac{1}{\sqrt 6}\tau_{11100}$\\
&& && $SD^{{(\pi N)} {(\rho N)}_3}_{31}$
&=& $-\frac{1}{2\sqrt 6}\tau_{11102} - \frac{1}{2\sqrt 2}\tau_{11202}$ \\
\hline
$\Delta_{3/2}^{-}$ && $K = 1, 2, 3$ &&
$D^{{(\pi N)} {(\eta \Delta)}}_{33}$
&=& $-\frac{1}{2}\tau_{22222}$ \\
&& && $D^{{(\pi N)} {(\pi N)}}_{33}$
&=& $\frac{1}{20}\tau_{11122} +\frac{1}{4}\tau_{22222}
+\frac{7}{10}\tau_{33322}$ \\
&& && $DS^{{(\pi N)} {(\pi \Delta)}}_{33}$
&=& $-\frac{1}{2\sqrt 5}\tau_{11120}$ \\
&& && $D^{{(\pi N)} {(\pi \Delta)}}_{33}$
&=& $\frac{\sqrt 2}{5\sqrt 5}\tau_{11122} +\frac{1}{\sqrt{10}}\tau_{22222}
-\frac{7}{5\sqrt{10}}\tau_{33322}$ \\
&& && $D^{{(\pi N)} {(\rho N)}_1}_{33}$
&=& $-\frac{1}{5\sqrt 6}\tau_{11122} -\frac{1}{10\sqrt 2}\tau_{11222}
-\frac{1}{2\sqrt{30}}\tau_{22122}$ \\
&& && && $-\frac{1}{3\sqrt 2}\tau_{22222}
-\frac{\sqrt 7}{6\sqrt{10}}\tau_{22322}$\\
&& && && $-\frac{7}{30\sqrt 2}\tau_{33222} -\frac{7}{15}\tau_{33322}$\\
&& && $DS^{{(\pi N)} {(\rho N)}_3}_{33}$
&=& $\frac{1}{4\sqrt 3}\tau_{11120} -\frac{\sqrt 5}{4\sqrt 3}\tau_{22120}$\\
&& && $D^{{(\pi N)} {(\rho N)}_3}_{33}$
&=& $-\frac{1}{20\sqrt 6}\tau_{11122}
-\frac{3}{20\sqrt 2}\tau_{11222} +\frac{1}{4\sqrt{30}}\tau_{22122}$ \\
&& && && $+\frac{1}{12\sqrt 2}\tau_{22222}
-\frac{\sqrt7}{3\sqrt{10}}\tau_{22322}$\\
&& && && $+\frac{7}{30\sqrt 2}\tau_{33222} +\frac{7}{15}\tau_{33322}$ \\
\hline
$\Delta_{5/2}^{-}$ && $K = 1, 2, 3$ &&
$D^{{(\pi N)} {(\eta \Delta)}}_{35}$
&=& $-\frac{\sqrt 7}{3\sqrt 2}\tau_{22222}$ \\
&& && $D^{{(\pi N)} {(\pi N)}}_{35}$
&=& $\frac{3}{10}\tau_{11122} +\frac{7}{18}\tau_{22222}
+\frac{14}{45}\tau_{33322}$ \\
&& && $D^{{(\pi N)} {(\pi \Delta)}}_{35}$
&=& $\frac{3\sqrt 7}{10\sqrt 5}\tau_{11122}
+\frac{\sqrt 7}{18\sqrt 5}\tau_{22222} -\frac{16\sqrt 7}{45\sqrt
5}\tau_{33322}$ \\
&& && $DG^{{(\pi N)} {(\pi \Delta)}}_{35}$
&=& $\frac{\sqrt 7}{3\sqrt{10}}\tau_{33324}$ \\
&& && $D^{{(\pi N)} {(\rho N)}_1}_{35}$
&=& $\frac{\sqrt 3}{10\sqrt 2}\tau_{11122} -\frac{1}{10\sqrt 2}\tau_{11222}
+\frac{7}{6\sqrt{30}}\tau_{22122}$ \\
&& && && $+\frac{7}{54\sqrt 2}\tau_{22222}
-\frac{\sqrt{14}}{27\sqrt 5}\tau_{22322}$\\
&& && && $+\frac{28\sqrt 2}{135}\tau_{33222} +\frac{7}{135}\tau_{33322}$\\
&& && $DG^{{(\pi N)} {(\rho N)}_3}_{35}$
&=& $-\frac{\sqrt 5}{9\sqrt 2}\tau_{22324} -\frac{\sqrt 7}{36}\tau_{33324}
-\frac{\sqrt{35}}{12\sqrt 3}\tau_{33424}$\\
&& && $D^{{(\pi N)} {(\rho N)}_3}_{35}$
&=& $\frac{\sqrt{21}}{20}\tau_{11122} -\frac{\sqrt 7}{20}\tau_{11222}
+\frac{\sqrt 7}{12\sqrt{15}}\tau_{22122}$ \\
&& && && $+\frac{13\sqrt 7}{108}\tau_{22222}
-\frac{17}{54\sqrt 5}\tau_{22322}$\\
&& && && $-\frac{2\sqrt 7}{135}\tau_{33222} +\frac{11\sqrt{14}}{135}\tau_{33322}$ \\
\hline
\end{tabular}
\end{table}

\section{Phenomenological Results} \label{phenom}

For the purpose of this analysis, we consider only 3- or 4-star
resonances as classified by the PDG~\cite{PDG}.  Only these
resonances are deemed to have unambiguous evidence of existence, and
moreover they tend to be the only ones for which BR into multiple
final states are available.  Since, for brevity, we truncate the
tables after spin $5/2$, unfortunately a few 3- and 4-star
resonances are missed in this analysis, namely, $N(2190) \, G_{17} \,
(N_{7/2}^-)$, $N(2220) \, H_{19} \, (N^+_{9/2})$, $N(2250)
\, G_{19} (N^-_{9/2})$, $N(2600) \, I_{1,11} \, (N^-_{11/2})$, $\Delta
(1950) \, F_{37} \, (\Delta^+_{7/2})$, and $\Delta (2420) \, H_{3,11}
(\Delta^+_{11/2})$.  The same approach described below applies to them
as well; however, for all of these cases the relative BR are poorly
known at best.

As we now show, the predictions of which resonances should be
associated with which poles---as determined by decay channels that
occur prominently versus those that are absent or weak---seem robust.
In particular, Eq.~(\ref{Mmaster}) can be employed in a
straightforward fashion to show that $\pi N \! \to \! \eta N$ contains
a single $K$ amplitude [with $K \! = \! L$], and the mixed partial
wave $\pi N (L) \! \to \pi \Delta (L^\prime)$ contains a single $K$
amplitude [with $K \! = \! \frac 1 2 (L \! + \!
L^\prime)$]~\cite{CLdecouple}.  For given $I$, $J$, and ${\cal P}$
these two amplitudes always probe distinct $K$, providing an
invaluable diagnostic.

However, once the reduced amplitude in which a pole occurs has been
determined, we also find that the prediction of the ratio of BRs
between two decay channels, as determined by the ratio of
Clebsch-Gordan coefficients (CGC) of the leading-order reduced
amplitudes in which the pole occurs, is not always in accord with
experiment.  Fortunately, these discrepancies can easily be explained
by $1/N_c$ corrections in the form of additional reduced amplitudes
occurring at that order, none of which is unnaturally large.  In
particular, there is enough information to predict the ratios of $\pi
N$ to $\pi \Delta$ BR's for a given resonance at leading [$O(N_c^0)$]
order, but most of them do not agree with experimental observation;
nevertheless, this effect is expected because it has been
shown~\cite{CDLN1} that transition amplitude relations for these
particular channels at leading order do not agree especially well with
experimental data, but the next-to-leading order relations work quite
well.  The following is our analysis for each channel.

\begin{enumerate}

\item $\bm{N^{+}_{1/2} \: (P_{11})}$:
The two well-established resonances in this channel are $N(1440)$ (the
Roper) and $N(1710)$.  In comparison, our transition amplitude
calculations provide two distinct pole structures, $K \! = \! 0$ and
$K \! = \! 1$.  It is certainly possible that the two known resonances
could be distinct poles in either the $K \! = \! 0$ or the 1
amplitudes, but the data suggests differently: The $N(1440)$ has a
very small, $(0 \! \pm \!  1) \%$, $\eta N$ BR while $N(1710)$ has a
small but nonnegligible $\eta N$ BR, $(6.2\pm 1.0)\%$.  The Roper
Breit-Wigner mass does in fact lie slightly below the $\eta N$
threshold (1485~MeV), but it is also a very broad state ($\Gamma$
perhaps as large as 450~MeV), making the total absence of an $\eta N$
final state noteworthy; indeed, some partial-wave analyses find the
even more kinematically suppressed $\rho N$ final state to occur at
the level of several percent.  Comparing this observation to our
tabulated result for the $\pi N \! \to \! \eta N $ transition
amplitude suggests that the Roper is a $K \! = \!  0$ pole and the
$N(1710)$ is a $K \! = \! 1$ pole.  This assignment also agrees very
well with the assumption of the Roper as a radial excitation of
ground-state $N$, which is a (nonresonant) $K \!  = \!  0$ state.
Unfortunately, the leading-order prediction of $\pi N \!
\to \! \pi N$ to $\pi N \! \to \! \pi \Delta$ BR's with this
assignment does not agree well with experiment.  For example, assuming
the $K \! = \! 0$ assignment, the $N(1440)$ $\pi N \! \to \! \pi
\Delta$ channel is predicted to have a BR $0.94$ times that of $\pi N
\! \to \! \pi N$ (the square of the relative CGC $\sqrt{2}$, reduced
by the smaller $\pi \Delta$ phase space), but experimentally this
ratio is only $0.38 \! \pm \!  0.10$.  The same is true for the
$N(1710)$ resonance but in the opposite fashion: The leading-order
prediction for the ratio of $\pi N \! \to \! \pi N$ to $\pi N \to \pi
\Delta$ BR for this resonance is $2.28$, but the experimental value is
$0.55 \! \pm \! 0.31$.  As mentioned above, this discrepancy can be
cured by 1/$N_c$-suppressed amplitudes.  To demonstrate this, we first
use Eq.~(2.6) of Ref.~\cite{CDLN2} to express the independent
$O(1/N_c)$ corrections---written in terms of $t$-channel
quantities---for both $\pi N \! \to \! \pi N$ and $\pi N \! \to \! \pi
\Delta$ channels: $-\frac{1}{N_c}\sqrt{\frac{2}{3}}[s^{t(+)}_{011} +
s^{t(-)}_{111}]$ and $-\frac{1}{N_c}\sqrt{\frac{1}{6}}[s^{t(+)}_{111}
+ s^{t(-)}_{211}]$, respectively.  (Using the result from
Ref.~\cite{CDLN2} that the independent $O(1/N_c^1)$ amplitudes have
$|I_t - \! J_t| \! = \! 1$, the notation is $s^{t(J_t \! - I_t)}_{I_t,
LL^\prime}$.)  Choosing the right combination of values for
$t$-channel reduced amplitudes $s^t$ carrying natural $O(1)$ values
that are not particularly fine-tuned, the experimentally observed
pattern of BR can easily be accommodated.  A similar result occurs for
the $\pi N \! \to \! \pi N$ to $\pi N \! \to \! \eta N$ ratio, for
which our leading-order prediction is $0.95$ and the experimental
value is $2.42 \! \pm \! 0.90$.  Also note that both $N(1440)$ and
$N(1710)$ are expected to have couplings to both $\rho N$ and $\omega
N$.  As noted above, hints of $N(1440) \! \to \! \rho N$ are seen in
some analysis, but $\omega N$ is not even suggested.  In fact, this
perhaps reflects the experimental difficulty of reconstructing the
$\omega$, which almost always contains a difficult-to-reconstruct
$\pi^0$ among its decay products.  Nevertheless, the PDG lists
$N(1710) \! \to \! \omega N$ with a BR of $(13.0 \!\pm \! 2.0) \%$,
alongside a $\rho N$ BR of 5--$25\%$; such appreciable branching
fractions are expected from our linear relations.  Indeed, the
relations predict separate results for $\rho N$ and $\omega N$ final
states for each distinct final spin configuration; should improved
data become available for such channels, the relations are ready to
confront them.

\item $\bm{N^{+}_{3/2} \: (P_{13})}$:  The only well-measured resonance
in this channel is $N(1720)$.  It has a small but nonnegligible $\eta
N$ BR, $(4.0 \! \pm \! 1.0)\%$, suggesting the assignment $K \! = \!
1$.  However, comparison of the $\pi N \! \to \! \eta N$ to $\pi N \!
\to \! \pi N$ leading-order prediction ($4.26$) to the experimental BR
($0.27 \! \pm \! 0.11$) contradicts this assignment.  Again, the
explanation may come from the $O(1/N_c^1)$ corrections, but in a
different way than discussed for the $N^+_{1/2}$ states: It is
possible that $N(1720)$ is actually a $K \! = \! 2$ pole and the $\eta
N$ BR comes purely from the $O(1/N_c)$ amplitudes.  The $N(1720)$
decays mostly to the $\rho N$ channel ($70$--$85\%$), in particular to
the $S^\prime=\frac{1}{2}$ $P$-wave, [one partial-wave analysis finds
a $(91 \! \pm \! 1)\%$ BR].  It has a large BR for this channel
despite the fact that $N(1720)$ has barely enough phase space for this
decay (threshold $\simeq 1708$~MeV), suggesting not only that some of
the reduced amplitudes are substantial, but also specific
cancellations among these amplitudes in other channels with much
smaller BR, such as $\pi N$ ($10$--$20\%$) or $\pi \Delta$ $P$-wave
(not listed as a separate BR but roughly comparable).  At this moment
neither $K \! = \! 1$ nor $K \! = \! 2$ is preferred; however, the
answer might be found in the mixed partial-wave $PF_{13}$ $\pi \Delta$
and $S^\prime \! = \! \frac 3 2$ $\omega N$ channels, where only a
single $K \! = \! 2$ amplitude appears.  Unfortunately, no data is yet
presented for these channels; indeed, the PDG inexplicably does not
even list the $F$-wave $\pi \Delta$ state as a possibility.

\item $\bm{N^{+}_{5/2} \: (F_{15})}$: Like the previous state,
only one well-measured resonance appears in this state, the $N(1680)$.
It has a very small, $(0.0 \! \pm \! 1.0)\%$ $\eta N$ BR, and a
considerable mixed wave $FP_{15}$ $\pi \Delta$ BR ($6$--$14\%$),
suggesting it to be a $K \!  = \! 2$ pole.  This resonance decays
mostly to the $\pi N$ channel ($65$--$70\%$), indicating the dominance
of the reduced amplitude $\tau_{22233}$ at the $K \! = \! 2$ pole.
This result appears to explain the considerable observed $\rho N$ BR
($3$--$15\%$) despite the very limited phase space available for this
decay, although there is not yet enough experimental information to
determine which of the three possible $\rho N$ channels is dominant,
nor whether the final-state $\omega N$ has ever been sought.  The
comparison of the ratio of the $FP_{15}$ $\pi \Delta$ to the unmixed
$F_{15}$ $\pi \Delta$ BR (the latter listed as $< \! 2\%$) compares
favorably with our prediction ($2.25$), but the $\pi N$ to $FP_{15}$
$\pi \Delta$ BR ratio is measured as $6.8 \! \pm \! 2.7$, versus our
prediction of 0.65, providing yet another example of the significance
of the $1/N_c$ corrections in such ratios, as well as the large size
of uncertainties in partial-wave analyses.

\item $\bm{N^{-}_{1/2} \: (S_{11})}$: The two prominent resonances in
this channel are $N(1535)$ and $N(1650)$.  Both resonances have
significant $\eta N$ BR, $(53 \! \pm \! 1)\%$ and $3$--$10\%$,
respectively, even though our leading-order results predict them to be
zero.  However, as was shown in the first work using this
method~\cite{CL1}, the $\eta N \! \to \! \eta N$ amplitude is purely
$K \! = \! 0$ at leading order, strongly suggesting that $N(1535)$ is
a $K \! = \! 0$ pole that has a $\pi N$ coupling through $O(1/N_c)$
mixing to $K \! = \! 1$, while $N(1650)$ is a $K \! = \! 1$ pole that
has an $\eta N$ coupling through $O(1/N_c)$ mixing to $K \! = \! 0$.
Further analysis for $\pi \Delta$ and $\rho N$ channels supports this
assignment.  For example, as was pointed out in
Ref.~\cite{CLdecouple}, the $K \! = \! 1$ $\pi \Delta$ mixed partial
wave $SD_{11}$ has a BR of $<1\%$ for $N(1535)$ but $1$--$7\%$ for
$N(1650)$.  Moreover, the $\rho$ and $\omega$ couplings are purely $K
\! = \!  1$ at leading order, while the $N(1535)$ has a $\rho N$ BR of
$<4\%$, the $N(1650)$ has $4$--$12\%$ (although available phase space
may be an important factor for these cases).

\item $\bm{N^{-}_{3/2} \: (D_{13})}$: The well-measured
resonances in this state are $N(1520)$ and $N(1700)$.  Both appear to
have appreciable mixed-wave $DS_{13}$ $\pi \Delta$ decay BR,
$5$--$12\%$ and (according to one analysis) $(11 \! \pm \! 1)\%$,
respectively, and essentially no $\eta N$ BR, $(0.23 \! \pm \!
0.04)\%$ and $(0 \! \pm \! 1)\%$, respectively, suggesting that both
are $K \! = \! 1$ poles.  However, they are stunningly different in
their $\pi N$ BR: 55--65$\%$ and 5--15$\%$, respectively.  One
possibility for explaining this confusing state of affairs is that the
$N(1700)$ $\eta N$ and $DS_{13}$ $\pi \Delta$ BR's, which come from
just one analysis, might be incorrect; indeed, the PDG lists the
$DS_{13}$ $\pi \Delta$ coupling as being consistent with zero, and an
older analysis with a larger $\eta N$ coupling is not used for the
PDG's estimate.  Under these circumstances, a $K \! = \! 2$ assignment
for $N(1700)$ certainly cannot be ruled out, in which case the $\pi N$
BR ratio between the two channels can be explained by the absolute
size of the $K \! = \! 2$ contribution to $\pi N \! \to \! \pi N$
relative to that from $K \! = \! 1$ being small.  $N(1520)$ has a
considerable $\rho N$ BR, $15$--$25\%$, even though it lies well below
the threshold given by $m_\rho \! + \!  m_N \! \simeq \! 1708$~MeV
(contrast this with comments for $N^{-}_{1/2}$), which can be
accommodated by the large value of $\tau_{11122}$ that contributes to
this channel and to $\pi N$.  The $DS_{13}$ and $D_{13}$ $\pi \Delta$
BR's for $N(1520)$ (the latter given as 10--14$\%$) are quite
comparable in magnitude; the experimental ratio is $0.71 \! \pm \!
0.31$, compared to our leading-order prediction of 2\@.  The $\pi N$
to $DS_{13}$ $\pi \Delta$ BR ratio for N(1520) is experimentally $7.1
\! \pm \! 3.0$, compared with the leading-order prediction of $0.76$;
again, the gulf seems large, but both the experimental uncertainties
and the effects of the $1/N_c$ corrections could serve to bridge the
gap.

\item $\bm{N^{-}_{5/2} \: (D_{15})}$: The only well-measured
resonance in this channel is $N(1675)$.  Its $\eta N$ BR is given as
$(0 \! \pm \! 1)\%$, suggesting that it is not a $K \! = \! 2$ pole.
However, the $DG_{15}$ $\pi \Delta$ amplitude, which is pure $K \! =
\! 3$ and therefore would resolve the issue, has apparently not been
extracted.  The $\pi N$ (35--45$\%$) and total $\pi \Delta$
(50--60$\%$) BR's are both substantial, but both contain both $K \! =
\! 2$ and $K \! = \! 3$ amplitudes.  Another mystery associated with
this resonance is its small $\rho N$ BR, $< 1$--$3\%$, despite
adequate phase space [especially compared to the $N(1520)$] and
couplings to both $K \! = \! 2$ and $K \! = \! 3$ amplitudes.  Without
further information, one cannot determine whether this effect is due
to a cancellation among the leading-order amplitudes or $1/N_c$
effects.  Clearly, determining the $DG_{15}$ $\pi \Delta$ BR is key to
resolving these issues.

\item $\bm{\Delta^{+}_{1/2} \: (P_{31})}$: The only well-measured
resonance in this channel is $\Delta(1910)$.  Unfortunately, not
enough experimental data exists to perform a meaningful analysis for
this state.  Since this particular channel does not admit a $\pi N \!
\to \pi \Delta$ mixed partial wave, in order to determine to which
pole this resonance belongs, one would need data for the $\eta \Delta$
BR (threshold $\simeq \!  1780$~MeV).  (In fact, no $\eta \Delta$
final state is tabulated for any $\Delta$ resonance.)  Knowing
only that the $\pi N$ BR is 15--30$\%$ does not narrow the
possibilities, since both $K \! = \! 1$ and $K \! = \! 2$ appear in
this channel.  In fact, more than one partial-wave analysis concludes
that a dominant decay channel of $\Delta (1910)$ is $\pi N(1440)$,
which if true cannot be handled in our current formalism since the
$N(1440)$ is not in the ground-state band.

\item $\bm{\Delta^{+}_{3/2} \: (P_{33})}$: The two prominent
resonances in this state are $\Delta(1600)$ and $\Delta(1920)$.  While
some analyses give evidence for the mixed partial wave $PF_{33}$ into
a $\pi \Delta$ final state for $\Delta (1600)$ [but not yet the
$\Delta(1920)$] indicative of a $K \! = \! 2$ pole, the numbers are
not sufficiently robust to determine the significance of the $PF_{33}$
channel, and hence the pole structure.  Indeed, the $\Delta (1600)$ is
often considered as a partner to the $N(1440)$, as a radial excitation
of the $\Delta(1232)$, in which case it would be a $K \! = \! 0$ pole;
the evidence for such an assignment would appear as and anomalously
small BR for both $\eta \Delta$ and $PF_{33}$ $\pi \Delta$.

\item $\bm{\Delta^{+}_{5/2} \: (F_{35})}$: The only prominent resonance
in this channel is $\Delta(1905)$.  The substantial $FP_{35}$ $\pi
\Delta$ BR ($23 \! \pm \! 1\%$, according to one analysis) indicates the
resonance to be a $K \! = \! 2$ pole.  The same analysis gives the
unmixed $\pi \Delta$ BR at $44 \! \pm \! 1\%$, for a ratio to $FP_{35}$ $\pi
\Delta$ of $1.91 \! \pm \! 0.09$ compared to the leading-order prediction
0.58.  The $\pi N$ BR is given as $12 \! \pm \! 3\%$, giving a ratio
to the $FP_{35}$ $\pi \Delta$ partial wave of $0.52 \! \pm \! 0.13$
vs.\ the leading-order prediction 0.113.  In fact, this resonance
appears to prefer $\rho \Delta$ decays ($>60\%$ by PDG estimate);
clearly, independent confirmation of the $\pi \Delta$ BR is necessary
to make a definitive assignment.

\item $\bm{\Delta^{-}_{1/2} \: (S_{31})}$: The only prominent resonance
in this channel is $\Delta(1620)$.  Since only $K \! = \! 1$
amplitudes appear at leading order in this case, it is natural to
assign $K \! = \! 1$ to the $\Delta (1620)$.  This resonance decays
largely (30--60$\%$) to the sole $\pi \Delta$ channel $SD_{31}$, and
only about 20--30$\%$ to $\pi N$ and 7--25$\%$ for $\rho N$.
According to one analysis, the $(\rho N)_1$ to $(\rho N)_3$ ratio of
BR's appears to be $7.0 \! \pm \! 3.8$ ($14 \! \pm \! 3\%$ to $2 \!
\pm \! 1\%$), but two distinct $K \! = \! 1$ amplitudes appear in the
expression for $(\rho N)_3$, making a direct comparison not yet
possible.  Most surprising, though, is the remarkably small $\pi N \!
\to \! \pi N$ BR, particularly considering that the same amplitude
appears in $\pi N \! \to \! (\rho N)_1$, and the former channel is
favored in BR only by about a factor of 2: The leading-order
prediction for the latter is suppressed not only by a CGC factor of 6,
but its phase space is also greatly suppressed (recall that $m_\rho \!
+ \!  m_N \! \simeq 1708$~MeV).  Again, $1/N_c$ corrections may
explain this huge discrepancy, but without confirmation of the BR it
is difficult to make such a sanguine prediction.

\item $\bm{\Delta^{-}_{3/2} \: (D_{33})}$: The only prominent
resonance in this channel is $\Delta(1700)$.  The $DS_{33}$ $\pi
\Delta$ partial wave actually appears to be the dominant mode
(25--50$\%$ BR), suggesting this resonance to be a $K \! = \! 1$ pole.
The unmixed ($D_{33}$) $\pi \Delta$ BR is only 1--7$\%$, while the
leading-order prediction BR ratio compared to $DS _{33}$ is 0.32, a
very favorable comparison.  $\pi N \! \to \! \pi N$ (10--20$\%$ BR) is
predicted at leading order to be only about 0.06 of the $DS_{33}$ $\pi
\Delta$ BR, which is about 1.5--3$\%$; again, resolving this discrepancy
is within the reach of $1/N_c$ corrections.  Too many amplitudes
appear in the each of the $\rho N$ channels (30--55$\%$ total BR) to
make any useful predictions.

\item $\bm{\Delta^{-}_{5/2} \: (D_{35})}$: The only prominent
resonance in this channel is $\Delta(1930)$.  Unfortunately there is
not sufficient experimental data to perform any meaningful analysis
for this state; indeed, although the $\pi N \! \to \! \pi N$ BR is
listed as only 5--15$\%$, one old analysis listed in the PDG states
that no $\pi \pi N$ final state was seen.  In light of the complicated
reduced amplitude structure for the $\Delta^-_{5/2}$, a determination
of the pole assignment of this resonance would require a measurement
of either the $\eta \Delta$ or $DG_{35}$ $\pi \Delta$ BR.

\end{enumerate}

The summary of our analysis is the pole assignments for the following
resonances: $K \! = \! 0$ are $N(1440)$ $P_{11}$ and $N(1535)$
$S_{11}$; $K \! = \! 1$ are $N(1710)$ $P_{11}$, $\Delta (1700)$
$P_{33}$, $N(1650)$ $S_{11}$, $N(1520)$ $D_{13}$, and $\Delta (1620)$
$S_{31}$; $K \! = \! 2$ are $N(1680)$ $F_{15}$ and $\Delta (1905)$
$F_{35}$.  Taking into account that the two values of ${\cal P}$ ($P$
and $F$ waves for ${\cal P} \! = \! +$, $S$ and $D$ waves for ${\cal
P} \! = \! -$) give separate multiplet structures, one concludes that
$N(1710)$ and $\Delta (1700)$ belong to a single $K^{\cal P} \! = \!
1^+$ multiplet, as do $N(1650)$, $N(1520)$, and $\Delta (1620)$
($K^{\cal P} \! = \! 1^-$), and $N(1680)$ and $\Delta (1905)$
($K^{\cal P} \! = \! 2^+$).  By the same token, $N(1440)$ is a $0^+$
and $N(1535)$ is a $0^-$.

Do the data support these as degenerate multiplets?  Certainly the
masses are consistent; the natural size of the mass splitting within a
degenerate multiplet is relatively $O(1/N_c^2)$ compared to the common
$O(N_c^1) \! \sim \! 1$~GeV baryon mass, of order 110~MeV or larger.
As demonstrated some time ago~\cite{JL}, even the $\Delta$-$N$ mass
difference $\delta \! = \! 290$~MeV is of a natural size in this
sense.  On the other hand, the range of well-known baryon resonances
is remarkably small: only from about 1.4 to 2.0~GeV\@.  The masses in
our proposed multiplets certainly pass this criterion.

There is also some evidence from the widths, which are $O(N_c^0)$ and,
if two resonances are in a multiplet, should be equal to about 1 part
in 3\@.  For the proposed $1^+$ multiplet they are 50--250 and
200--400~MeV, respectively; for the $1^-$ they are 145--185, 100--125,
and 135--150~MeV, respectively; and for the $2^+$ multiplet they are
120--140 and 270--400~MeV.  Only the last of these should warrant
attention as a possible problem, and even in this case the large
difference in width may simply be attributable to the large number of
final-state channels that open between 1680 and 1905~MeV.

Those resonances for which we conclude pole assignments are not
unambiguous (but could be made so by observing $\eta$ final states or
$\pi \Delta$ mixed partial waves) are $N(1675)$ $D_{15}$ ($K \! = \!
2$ or 3), $N(1700)$ $D_{13}$ ($K \! = \! 1$ or 2), $N(1720)$ $P_{13}$
($K \! = \! 1$ or 2), $\Delta (1600)$ and $\Delta (1920)$ $P_{33}$ ($K
\! = \! 0, 1$, or 2), $\Delta (1910)$ $P_{31}$ ($K \! = \! 1$ or 2),
and $\Delta (1930)$ $D_{35}$ ($K \! = \! 1, 2$, or 3).  While this
level of ambiguity may be bothersome, we hasten to point out the
multitude of opportunities for tests.  For example, the $K^{\cal P} \!
= \! 2^-$ pole appears in no less than 28 of our observable partial
waves.

\section{Discussion and Conclusions} \label{concl}

As should be clear from this analysis, the constraints provided by the
$1/N_c$ expansion seem reliable in most cases for predicting the
identity of a resonant pole; the states with a final $\eta$ and the
mixed partial-wave $\pi \Delta$ final states, each being sensitive to
a single $K$ amplitude, make this possible.  (It should be added that
the $\pi \! \to \! \omega$ mixed partial-wave amplitudes, if they are
ever measured, share this property.) However, the particular coupling
ratios given at leading order by the group-theoretical constraints of
the $1/N_c$ expansion do not fare as well, even when phase space
corrections are included.  We have referred several times to the need
for including $1/N_c$ corrections in order to explain these large
discrepancies with data, and in the first example ($N^+_{1/2}$)
presented the explicit expressions for the $1/N_c$-suppressed
amplitudes to show that corrections of the required size are in fact
possible.  However, the true measure of whether the $1/N_c$ expansion
is in genuine agreement with the full data set would require a global
fit to the multitude of physical scattering amplitudes, many of which
are poorly determined or have never been determined at all.

In comparison, the older works of Refs.~\cite{CGKM,CC,GSS,GS} employ
specific operators at both leading and subleading orders in $1/N_c$,
to consider $\eta N$ and $\Delta N$ final states, their starting point
being well-defined spin-flavor multiplets such as [SU(6),$\, J^P$] =
({\bf 70}, $1^-$).  As discussed in great detail starting in
Refs.~\cite{CL1,CLcompat}, this ``operator'' or ``Hamiltonian''
approach is absolutely legitimate when either the baryons are stable
against strong decays or are comprised of heavy valence quarks, in
which case the eigenstates of a Hamiltonian consisting of operators
with well-defined spin-flavor transformation properties are the
asymptotically free states of the theory.  Indeed, such analyses were
patterned after works such as Refs.~\cite{DJM,JL} and were introduced
for the excited baryons in Refs.~\cite{CGKM,CCGL}.  It is not clear in
this strict approach, however, that substantial baryon resonance
widths or configuration mixing between distinct spin-flavor multiplets
can be accommodated, which is why the present ``scattering'' approach
was developed.  The old spin-flavor multiplets of SU(6)$\, \times
\,$O(3) were shown in Ref.~\cite{CLcompat} to be compatible with the
scattering approach, meaning that they form reducible collections of
the true large $N_c$ irreducible degenerate resonance multiplets.  The
numerical results of Refs.~\cite{CGKM,CC,GSS,GS} are certainly have
smaller stated uncertainties than the ones in this work (ours being
presented only at leading order in $N_c$), but this reflects the fact
that their formalism is more limited by considering only particular
spin-flavor multiplets, which are effectively assumed to be narrow in
width and unmixed with other states.  As mentioned above, it is
possible in principle to perform a global analysis of all baryon
resonances in our formalism, including $1/N_c$ corrections.  Finally,
we point out that the older works did not consider final-state vector
mesons.

In the context of our formalism, how might a violation of the $1/N_c$
expansion manifest itself?  Let us suppose that the analysis of the
data, including the large data set being collected at Jefferson Lab,
extracts branching ratios with smaller uncertainties for all the
expected partial waves.  A large branching ratio for a given resonance
into both the $\eta$ and mixed partial-wave $\pi \Delta$ final state,
corresponding to two distinct values of $K$, would be problematic for
this program.  Alternately, a partial wave that gives no signal of a
pole determined through other channels but nevertheless contains the
requisite reduced amplitudes [for example, if a prominent pole seen in
$D_{15}^{(\pi N)(\eta N)}$, giving $K^{\cal P} \! = \! 2^-$, did not
appear in $D_{15}^{(\pi N)(\pi N)}$] would spell trouble for the
approach.  Finally, a global fit of the sort described above, in which
some of the reduced amplitudes turned out to be substantially larger
than unity, would indicate a problem.

As we discussed in the Introduction, the baryon resonances do indeed
represent a murky corner of QCD; however, the $1/N_c$ expansion
already provides a bit of clarity, and promises to provide much more
in the future.

\section*{Acknowledgments}
The authors thank Tom Cohen for valuable discussions.  This work was
supported by the NSF under Grant No.\ PHY-0456520.

\end{document}